\documentclass[conference]{IEEEtran}
\PassOptionsToPackage{ruled, linesnumbered}{algorithm2e}
\usepackage[T1]{fontenc}
\usepackage[utf8]{inputenc}
\usepackage{mathrsfs}
\usepackage{algorithm2e}
\usepackage{amsmath}
\usepackage{amssymb}
\usepackage{graphicx}
\usepackage[bookmarks=true,bookmarksnumbered=true,bookmarksopen=true,bookmarksopenlevel=1,
 breaklinks=false,pdfborder={0 0 0},pdfborderstyle={},backref=false,colorlinks=false]
 {hyperref}
\hypersetup{pdftitle={Your Title},
 pdfauthor={Your Name},
 pdfpagelayout=OneColumn, pdfnewwindow=true, pdfstartview=XYZ, plainpages=false}

\makeatletter
\usepackage[caption=false,font=footnotesize]{subfig}
\usepackage{amsmath, amssymb, amsfonts, graphicx}

\allowdisplaybreaks
\newcommand{\herm}{^{\mathsf{H}}}
\newcommand{\trans}{^{\mathsf{T}}}

\DeclareMathOperator{\minimize}{minimize}
\DeclareMathOperator{\diag}{diag}

\DeclareMathOperator{\st}{subject~to}

\makeatother

\begin{document}
\title{Green Integration of Sensing, Communication, and Power Transfer via STAR-RIS}
\author{\IEEEauthorblockN{Vaibhav~Kumar\IEEEauthorrefmark{1}, and Marwa~Chafii\IEEEauthorrefmark{1}\IEEEauthorrefmark{2}}\IEEEauthorblockA{\IEEEauthorrefmark{1}Engineering Division, New York University Abu Dhabi, UAE\\
\IEEEauthorrefmark{2}NYU WIRELESS, NYU Tandon School of Engineering, New York, USA \\
Email: vaibhav.kumar@ieee.org, marwa.chafii@nyu.edu}}
\maketitle
{\let\thefootnote\relax\footnotetext{This work was supported in part by the Center for Cybersecurity under the New York University Abu Dhabi Research Institute under Award G1104.}}
\begin{abstract}
The upcoming sixth-generation (6G) wireless standard is anticipated to support a variety of applications that will require a seamless integration of sensing and communication services in a single network infrastructure. At the same time, 6G is also expected to support millions of low-powered Internet-of-Things (IoT) devices. Previous studies have demonstrated that the power requirements of these IoT devices can be met through wireless power transfer facilitated by intelligent metasurfaces. Therefore, in this paper, we try to formulate and answer a fundamental question: How much transmit power is required for an integrated sensing, communication, and power transfer (ISCPT) system? More specifically, we consider the problem of optimal active, passive, and receive beamforming design for a simultaneously transmitting and reflecting reconfigurable intelligent surface (STAR-RIS)-enabled ISCPT system with multiple sensing targets, multiple information receivers, and multiple energy receivers, to minimize the required transmit power from the base station, while guaranteeing predefined sensing, communication, and energy harvesting requirements. To tackle the challenging non-convex optimization problem, we use an alternating optimization (AO)-based approach, where the receive beamforming is obtained in closed form while the active and passive beamforming are obtained using a second-order cone program (SOCP) approach. Our numerical results show the benefit of using STAR-RIS to reduce the transmit power requirement for the ISCPT system compared to its corresponding conventional RIS (cRIS)-enabled and non-RIS ISCPT systems. 
\end{abstract}


\IEEEpeerreviewmaketitle{}

\section{Introduction\protect\label{sec:Introduction}}

Due to the various emerging applications, e.g., autonomous driving, smart city, unmanned aerial vehicles (UAVs), remote health monitoring, smart manufacturing, and industrial Internet-of-Things (IoT), the sixth generation (6G) wireless communication standard is being foreseen as a big paradigm shift from its predecessors. Supporting these services will inherently require a seamless integration of communication and sensing services, which are offered as separate functionalities in the existing wireless standards~\cite{24-WC-network_ISAC}. Therefore, integrated sensing and communication (ISAC) has gained tremendous attention in recent years as a technology enabler for 6G~\cite{23-SPM-separation_integration}. It has also been well-established that sensing and communication functionalities can achieve mutual benefit from each other when operated in a collaborative fashion, while an intrinsic tradeoff exists when they operate in a competitive manner~\cite{23-TWC-detectionProbability_achievableRate}. Using a mutual-information framework, the superiority of an ISAC system over a frequency-division sensing-and-communication (FDSAC) system in terms of sensing-communication (Se-Co) rate region was shown in~\cite{23-ComMag-MI_ISAC}, where the former had a larger Se-Co regime compared to the latter. 

Interestingly, wireless signals not only carry information but also contain energy, which has driven significant interest in wireless power transfer research over the past decades. More recently, integrating sensing, communication, and wireless power transfer (ISCPT) has became a prominent topic in signal processing. In this direction, a fundamental overview, a discussion on theoretical performance limits, and a practical implementation of ISCPT system was presented in~\cite{24-ComMag-ISCPT}. A comprehensive study on the performance trade-off limits among sensing, communication, and power transfer in an ISCPT system was presented in~\cite{24-TWC-ISAC-meets-SWIPT}, where the performance was analyzed in terms of the estimation Cram\'er-Rao bound (CRB), communication rate, and harvested power. Along similar lines, the problem of beamforming design in an ISCPT system to optimize the sensing performance while guaranteeing communication and power transfer quality-of-services (QoSs) was presented in~\cite{24-JSAC-ISCPT-Beamforming}. These existing studies confirm that ISCPT systems results in notable superior performance over time-division-based and eigenmode transmission (EMT)-based designs. 

At the same time, intelligent metasurfaces have been well-established as a groundbreaking hardware technology to enhance the performance of next-generation wireless systems. A detailed survey on recent trends, challenges, and open questions on intelligent metasurface-aided ISAC was presented in~\cite{2024_Magbool}. In~\cite{EA-TWC-MultiTarget-STARS-ISAC}, the authors devised a beamforming design solution to a STAR-RIS-enabled ISAC system with multiple communication users and targets. In particular, a signature sequence modulation scheme was introduced therein to enable multi-target detection at the base station, and then a semi-definite programming (SDP)-based beamforming solution was provided to maximize the minimum beampattern gain at the targets. Furthermore, an efficient beamforming design for STAR-RIS-enabled bi-static ISAC under high mobility settings was proposed in~\cite{24-TCOM-HighMobility-STARS-ISAC}. 

Contrary to the existing literature on metasurface-assisted ISAC, the literature on metasurface-assisted ISCPT is very limited. To the best of our knowledge, the answer to one of the most fundamental questions of how much transmit power is required to integrate sensing, communication, and power transfer in a single system, is still unknown. Against this background, our main contributions of this paper are as follows:
\begin{enumerate}
\item[1)]  We formulate a transmit power minimization problem for a STAR-RIS-enabled ISCPT system to guarantee a predefined sensing, communication, and power transfer constraints. To solve the challenging non-convex problem, we adopt an alternating optimization (AO)-based approach, where the receive beamforming is obtained in closed-form, while the active and passive beamforming are obtained via a successive convex approximation (SCA)-based approach. 
\item[2)]  For the proposed AO-based algorithm, we present a convergence analysis, and also obtain the associated computational complexity. 
\item[3)]  We then perform extensive numerical experiments to evaluate the performance of the proposed STAR-RIS-enabled system, and also to showcase the impact of different system design parameters on its performance. Our results establish the benefit and superiority of the proposed system over some of the benchmarks in terms of reducing the total required transmit power in the integrated system. The results also show that STAR-RIS is highly effective in extending the operating distance of the energy receivers from the transmitter and STAR-RIS. 
\end{enumerate}

\paragraph*{Notation}

Boldface uppercase and lowercase letters denote matrices and vectors, respectively. Using $\mathbb{C}^{M\times N}$/$\mathbb{R}^{M\times N}$, we denote the vector space of all complex-valued/real-valued matrices of size $M\times N$. $(\cdot)\trans$, $(\cdot)\herm$, $|\cdot|$, and $\|\cdot\|$ represent the (ordinary) transpose, conjugate transpose, absolute value, and Frobenius norm, respectively. $\Re\{\mathbf{X}\}$ and $\Im\{\mathbf{X}\}$ respectively denote the real and imaginary components of the complex-valued matrix $\mathbf{X}$. $j$ represents $\sqrt{-1}$, and $\mathbf{X}^{(\varepsilon)}$ represents the value of the optimization variable $\mathbf{X}$ at the $\varepsilon$-th iteration. $\mathcal{O}(\cdot)$ represents the asymptotic notation. 

\section{System Model and Problem Formulation}

\begin{figure}[tb]
\begin{centering}
\includegraphics[width=0.95\columnwidth]{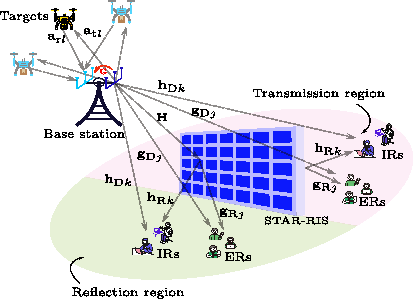}
\par\end{centering}
\caption{System model for STAR-RIS-enabled ISCPT.}
\label{fig:system_model}
\end{figure}

We consider the STAR-RIS-enabled ISCPT system shown in~Fig.~\ref{fig:system_model}, consisting of a dual-function radar-communication base station (B), one STAR-RIS (S), $K_{\mathrm{I}}$ information receivers (IRs), $K_{\mathrm{E}}$ energy receivers (ERs), and $L$ passive targets. It is assumed that B is equipped with $N_{\mathrm{t}}$ transmit antennas and $N_{\mathrm{r}}$ receive antennas, S is equipped with $N_{\mathrm{s}}$ simultaneously transmitting and reflecting elements, while the IRs and ERs are all single-antenna nodes. By $\mathbf{H}\in\mathbb{C}^{N_{\mathrm{s}}\times N_{\mathrm{t}}}$, $\mathbf{h}_{\mathrm{D}k}\in\mathbb{C}^{1\times N_{\mathrm{t}}}$, and $\mathbf{g}_{\mathrm{D}\jmath}\in\mathbb{C}^{1\times N_{\mathrm{t}}}$, we denote the wireless channel between B and S, B and $k$-th IR ($\forall k\in\mathcal{K}_{\mathrm{I}}\triangleq\{1,2,\ldots,K_{\mathrm{I}}\}$), and B and $\jmath$-th ER ($\forall\jmath\in\mathcal{K}_{\mathrm{E}}\triangleq\{1,2,\ldots,K_{\mathrm{E}}\}$), respectively. At the same time, the wireless channel for the links between S and $k$-th IR and S and $\jmath$-th ER are respectively denoted by $\mathbf{h}_{\mathrm{R}k}\in\mathbb{C}^{1\times N_{\mathrm{s}}}$ and $\mathbf{g}_{\mathrm{R}\jmath}\in\mathbb{C}^{1\times N_{\mathrm{s}}}$. Moreover, $\mathbf{a}_{\mathrm{t}l}(\phi_{l})\in\mathbb{C}^{1\times N_{\mathrm{t}}}$ and $\mathbf{a}_{\mathrm{r}l}(\phi_{l})\in\mathbb{C}^{N_{\mathrm{r}}\times1}$ respectively denotes the forward and backward channels between B and $l$-th target, with $\phi_{l}$ being the azimuth angle of the $l$-th target ($\forall l\in\mathcal{L}\triangleq\{1,2,\ldots,L\}$). The self-interference channel between the transmit and receive antenna arrays at B is denoted by $\mathbf{G}\in\mathbb{C}^{N_{\mathrm{r}}\times N_{\mathrm{t}}}$. We assume that the links between STAR-RIS and targets are blocked due to severe path loss and/or blockage. 

Note that in the system under consideration, the IRs are interested in the information/message transmitted from B, while the ERs are only interested in harvesting the energy contained in the broadcast signal from B. Hence B broadcasts a linear superposition of sensing, communication and energy harvesting signals as follows: 
\begin{equation}
\mathbf{x}=\sum\nolimits_{l\in\mathcal{L}}\mathbf{f}_{\mathrm{T}l}w_{\mathrm{T}l}+\sum\nolimits_{k\in\mathcal{K}_{\mathrm{I}}}\mathbf{f}_{\mathrm{I}k}w_{\mathrm{I}k}+\sum\nolimits_{\jmath\in\mathcal{K}_{\mathrm{E}}}\mathbf{f}_{\mathrm{E}\jmath}w_{\mathrm{E}\jmath},\label{eq:transmit_signal}
\end{equation}
 where $w_{\mathrm{T}l}$, $w_{\mathrm{I}k}$, and $w_{\mathrm{E}\jmath}$ denote the $l$-th target signal, $k$-th communication signal, and $\jmath$-th energy harvesting signal, while $\mathbf{f}_{\mathrm{T}l},\mathbf{f}_{\mathrm{I}k},\mathbf{f}_{\mathrm{E}\jmath}\in\mathbb{C}^{N_{\mathrm{t}}\times1}$ are the corresponding beamforming vectors. For the simplicity of analysis, we use a \emph{slight abuse of notation} to define $\mathbf{F}\triangleq[\mathbf{f}_{\mathrm{T}1},\ldots,\mathbf{f}_{\mathrm{T}L},\mathbf{f}_{\mathrm{I}1},\ldots,\mathbf{f}_{\mathrm{I}K_{\mathrm{I}}},\mathbf{f}_{\mathrm{E}1},\ldots,\mathbf{f}_{\mathrm{E}K_{\mathrm{E}}}]=[\mathbf{f}_{1},\ldots,\mathbf{f}_{L},\mathbf{f}_{L+1},\ldots,\mathbf{f}_{L+K_{\mathrm{I}}},\mathbf{f}_{L+K_{\mathrm{I}}+1},\ldots,\mathbf{f}_{L+K_{\mathrm{I}}+K_{\mathrm{E}}}]\in\mathbb{C}^{N_{\mathrm{t}}\times(L+K_{\mathrm{I}}+K_{\mathrm{E}})}$ as the stack of all the active beamforming vectors. In this paper, we consider the power-splitting (PS) model of the STAR-RIS, i.e., for an incident signal $x$, the response of the $\varkappa$-th STAR-RIS element in the reflection and transmission regimes are respectively given by $\theta_{\mathsf{R}\varkappa}x$ and $\theta_{\mathsf{T}\varkappa}x$, provided $\angle\theta_{\mathsf{R}\varkappa},\angle\theta_{\mathsf{T}\varkappa}\in[0,2\pi)$ and $|\theta_{\mathsf{R},\varkappa}|^{2}+|\theta_{\mathsf{T},\varkappa}|^{2}=1\ \forall\varkappa\in\mathscr{N}_{\mathrm{s}}\triangleq\{1,2,\ldots,N_{\mathrm{s}}\}$.\footnote{Although a more practical coupled transmission and reflection phase-shift model was recently proposed for STARS in~\cite{22-JSAC-CorrelatedModel}, in this paper we stick to the widely used non-coupled phase-shift model for the sake of tractability.} The received signal at $k$-th IR is then given by 
\begin{equation}
y_{\mathrm{I}k}=(\mathbf{h}_{\mathrm{D}k}+\mathbf{h}_{\mathrm{R}k}\boldsymbol{\Theta}_{\mathrm{I}k}\mathbf{H})\mathbf{x}+n_{\mathrm{I}k}=\mathbf{z}_{\mathrm{I}k}\mathbf{x}+n_{\mathrm{I}k},\label{eq:received_signal_Ik}
\end{equation}
 where $\mathbf{z}_{\mathrm{I}k}\triangleq\mathbf{h}_{\mathrm{D}k}+\mathbf{h}_{\mathrm{R}k}\boldsymbol{\Theta}_{\mathrm{I}k}\mathbf{H}$, $n_{\mathrm{I}k}\sim\mathcal{CN}(0,\sigma_{\mathrm{I}k}^{2})$ is the additive white Gaussian noise (AWGN) at the $k$-th IR, $\boldsymbol{\Theta}_{\mathrm{I}k}=\diag(\boldsymbol{\theta}_{\mathrm{I}k})$ is the passive beamforming for the $k$-th IR. Moreover, $\boldsymbol{\theta}_{\mathrm{I}k}=\boldsymbol{\theta}_{\mathsf{R}}=[\theta_{\mathsf{R}1},\ldots,\theta_{\mathsf{R}N_{\mathrm{s}}}]\trans$ if $k\in\mathcal{K}_{\mathrm{I},\mathsf{R}}$ and $\boldsymbol{\theta}_{\mathrm{I}k}=\boldsymbol{\theta}_{\mathsf{T}}=[\theta_{\mathsf{T}1},\ldots,\theta_{\mathsf{T}N_{\mathrm{s}}}]\trans$ if $k\in\mathcal{K}_{\mathrm{I},\mathsf{T}}$; here $\mathcal{K}_{\mathrm{I},\mathsf{R}}$ and $\mathcal{K}_{\mathrm{I},\mathsf{T}}$ denote the indices of the IRs in the reflection and transmission regimes of the STAR-RIS, respectively, provided $\mathcal{K}_{\mathrm{I},\mathsf{R}}\cap\mathcal{K}_{\mathrm{I},\mathsf{T}}=\emptyset$ and $\mathcal{K}_{\mathrm{I},\mathsf{R}}\cup\mathcal{K}_{\mathrm{I},\mathsf{T}}=\mathcal{K}_{\mathrm{I}}$. Hence, we can model the signal-to-interference-plus-noise ratio (SINR) at the $k$-th IR to decode $w_{\mathrm{I}k}$ by 
\begin{equation}
\gamma_{\mathrm{I}k}(\mathbf{F},\boldsymbol{\theta})=\frac{|\mathbf{z}_{\mathrm{I}k}\mathbf{f}_{L+k}|^{2}}{\sum\nolimits_{\imath\in\mathcal{Q}\setminus\{L+k\}}|\mathbf{z}_{\mathrm{I}k}\mathbf{f}_{\imath}|^{2}+\sigma_{\mathrm{I}k}^{2}},\label{eq:def_SINR_Ik}
\end{equation}
where $\boldsymbol{\theta}\triangleq[\boldsymbol{\theta}_{\mathsf{R}}\trans,\boldsymbol{\theta}_{\mathsf{T}}\trans]\trans$ and $\mathcal{Q}\triangleq\{1,2,\ldots,L+K_{\mathrm{I}}+K_{\mathrm{E}}\}$. Similarly, the harvested power at the $\jmath$-th ER can be modeled as 
\begin{equation}
p_{\mathrm{E}\jmath}(\mathbf{F},\boldsymbol{\theta})=\eta\sum\nolimits_{\imath\in\mathcal{Q}}|\mathbf{z}_{\mathrm{E}\jmath}\mathbf{f}_{\imath}|^{2},\label{eq:def_harvested_power_Ej}
\end{equation}
where $0<\eta<1$ is the power harvesting efficiency for the $\jmath$-th ER ($\forall\jmath\in\mathcal{K}_{\mathrm{E}}$), $\mathbf{z}_{\mathrm{E}\jmath}\triangleq\mathbf{g}_{\mathrm{D}\jmath}+\mathbf{g}_{\mathrm{R}\jmath}\boldsymbol{\Theta}_{\mathrm{E}\jmath}\mathbf{H}$, and $\boldsymbol{\Theta}_{\mathrm{E}\jmath}=\diag(\boldsymbol{\theta}_{\mathrm{E}\jmath})$. Moreover, $\boldsymbol{\theta}_{\mathrm{E}\jmath}=\boldsymbol{\theta}_{\mathsf{R}}=[\theta_{\mathsf{R}1},\ldots,\theta_{\mathsf{R}N_{\mathrm{s}}}]\trans$ if $\jmath\in\mathcal{K}_{\mathrm{E},\mathsf{R}}$ and $\boldsymbol{\theta}_{\mathrm{E}\jmath}=\boldsymbol{\theta}_{\mathsf{T}}=[\theta_{\mathsf{T}1},\ldots,\theta_{\mathsf{T}N_{\mathrm{s}}}]\trans$ if $\jmath\in\mathcal{K}_{\mathrm{E},\mathsf{T}}$; here $\mathcal{K}_{\mathrm{E},\mathsf{R}}$ and $\mathcal{K}_{\mathrm{E},\mathsf{T}}$ denote the indices of the ERs in the reflection and transmission regimes of the STAR-RIS, respectively, provided $\mathcal{K}_{\mathrm{E},\mathsf{R}}\cap\mathcal{K}_{\mathrm{E},\mathsf{T}}=\emptyset$ and $\mathcal{K}_{\mathrm{E},\mathsf{R}}\cup\mathcal{K}_{\mathrm{E},\mathsf{T}}=\mathcal{K}_{\mathrm{E}}$. At the same time, the echo signal from the targets to the receive array of B is given by 
\begin{equation}
\mathbf{y}_{\mathrm{B}}=\sum\nolimits_{l}\alpha_{l}\mathbf{A}_{l}(\phi_{l})\mathbf{x}+\mathbf{G}\mathbf{x}+\mathbf{n}_{\mathrm{B}},\label{eq:received_echo_signal}
\end{equation}
 where $\alpha_{l}$ corresponds to the $l$-th target's radar cross-section (RCS), $\mathbf{A}_{l}(\phi_{l})=\mathbf{a}_{\mathrm{r}l}(\phi_{l})\mathbf{a}_{\mathrm{t}l}(\phi_{l})$, and $\mathbf{n}_{\mathrm{B}}\sim\mathcal{CN}(\boldsymbol{0},\sigma_{\mathrm{B}}^{2}\mathbf{I})$ is the AWGN vector at the receive array of B. Before processing the received echo signal, B applies the receive beamforming vector $\mathbf{u}_{l}\in\mathbb{C}^{N_{\mathrm{r}}\times1}$ to sense the $l$-th target; this results in the following sensing SINR for the $l$-th target
\begin{align}
 & \gamma_{\mathrm{T}l}(\mathbf{F},\mathbf{U})=\bar{\alpha}_{l}\sum\nolimits_{\imath\in\mathcal{Q}}|\mathbf{u}_{l}\herm\mathbf{A}_{l}(\phi_{l})\mathbf{f}_{\imath}|^{2}\Big/\Big(\sum\nolimits_{\imath\in\mathcal{Q}}|\mathbf{u}_{l}\herm\mathbf{G}\mathbf{f}_{\imath}|^{2}\nonumber \\
 & \quad+\sum\nolimits_{\ell\in\mathcal{L}\setminus\{l\}}\bar{\alpha}_{\ell}\sum\nolimits_{\imath\in\mathcal{Q}}|\mathbf{u}_{l}\herm\mathbf{A}_{\ell}(\phi_{\ell})\mathbf{f}_{\imath}|^{2}+\sigma_{\mathrm{B}}^{2}\|\mathbf{u}_{l}\|^{2}\Big),\label{eq:def_echo_SINR_Sl}
\end{align}
where $\mathbf{U}=[\mathbf{u}_{1},\ldots,\mathbf{u}_{L}]\in\mathbb{C}^{N_{\mathrm{r}}\times L}$ is the stack of receive beamforming vectors, and $\bar{\alpha}_{l}=\mathbb{E}\{|\alpha_{l}|^{2}\}\ \forall l\in\mathcal{L}$.\footnote{Here we assume that all the targets are resolvable at the base station; the details of the underlying signal processing algorithm is beyond the scope of this paper.} 

With the given background, the problem of transmit power minimization from B can be formulated as follows:
\begin{subequations}
\label{eq:P1}
\begin{eqnarray}
(\mathbb{P}1)\  & \underset{\mathbf{F},\boldsymbol{\theta},\mathbf{U}}{\minimize}\  & \|\mathbf{F}\|^{2},\label{eq:P1-objective}\\
 & \st & \gamma_{\mathrm{I}k}(\mathbf{F},\boldsymbol{\theta})\geq\Gamma_{\mathrm{I}k}\ \forall k\in\mathcal{K}_{\mathrm{I}},\label{eq:P1-comm-QoS}\\
 &  & \frac{1}{\Delta_{\jmath}}p_{\mathrm{E}\jmath}(\mathbf{F},\boldsymbol{\theta})\geq1\ \forall\jmath\in\mathcal{K}_{\mathrm{E}},\label{eq:P1-energy-harvesting-QoS}\\
 &  & \gamma_{\mathrm{T}l}(\mathbf{F},\mathbf{U})\geq\Gamma_{\mathrm{T}l}\ \forall l\in\mathcal{L},\label{eq:P1-sensing-QoS}\\
 &  & \|\mathbf{u}_{l}\|^{2}=1\ \forall l\in\mathcal{L},\label{eq:UMC-receive-beamforming}\\
 &  & |\theta_{\mathsf{R},\varkappa}|^{2}+|\theta_{\mathsf{T},\varkappa}|^{2}=1\ \forall\varkappa\in\mathscr{N}_{\mathrm{s}}.\label{eq:P1-UMC}
\end{eqnarray}
\end{subequations}
In~($\mathbb{P}1$),~\eqref{eq:P1-objective} represents the total transmit power from B,~\eqref{eq:P1-comm-QoS} enforces communication QoS for IRs where $\Gamma_{\mathrm{I}k}$ is the communication SINR threshold,~\eqref{eq:P1-energy-harvesting-QoS} ensures that the harvested power at $\jmath$-th ER is equal to or greater than the required threshold $\Delta_{\jmath}$,~\eqref{eq:P1-sensing-QoS} constraints the sensing QoS with $\Gamma_{\mathrm{T}l}$ being the echo SINR threshold for $l$-th target,~\eqref{eq:UMC-receive-beamforming} ensures that the received echo signal is not amplified during receive beamforming, and~\eqref{eq:P1-UMC} is due to the PS mode of the STAR-RIS. 

One can easily note that~($\mathbb{P}1$) is non-convex due to the coupling between the design variables in~\eqref{eq:P1-comm-QoS},~\eqref{eq:P1-energy-harvesting-QoS}, and~\eqref{eq:P1-sensing-QoS}, and also due to the equality constraint in~\eqref{eq:P1-UMC}. Therefore, obtaining a stationary solution to~($\mathbb{P}1$) is very challenging. 

\section{Proposed Solution}

In this section, we propose an efficient numerical solution to~($\mathbb{P}1$) using AO and convex approximations. In particular, we first obtain the optimal receive beamforming vectors, i.e., $\mathbf{U}$ in closed-form while keeping the active beamforming vectors ($\mathbf{F}$) and passive beamforming vectors ($\boldsymbol{\theta}$) fixed. Next, for the given optimal $\mathbf{U}$, we use a series of convex approximations to update $\mathbf{F}$ and $\boldsymbol{\theta}$. 

\subsection{Optimizing the Receive Beamforming Vectors }

Using~\eqref{eq:def_echo_SINR_Sl}, and the notion of generalized Rayleigh quotient, the optimal receive beamforming vector $\mathbf{u}_{l,\mathrm{opt}}$ for a fixed $\mathbf{F}$ can be given by 
\begin{equation}
\mathbf{u}_{l,\mathrm{opt}}=\boldsymbol{\lambda}_{\max}\big\{\mathbf{M}_{2l}^{-1}\mathbf{M}_{1l}\big\},\label{eq:optimal-receive-beamforming-closed}
\end{equation}
where $\mathbf{M}_{1l}=\bar{\alpha}_{l}\mathbf{A}_{l}(\phi_{l})\mathbf{F}\mathbf{F}\herm\mathbf{A}_{l}\herm(\phi_{l})$, $\mathbf{M}_{2l}=\mathbf{G}\mathbf{F}\mathbf{F}\herm\mathbf{G}\herm+\sum\nolimits_{\ell\in\mathcal{L}\setminus\{l\}}\bar{\alpha}_{\ell}\mathbf{A}_{\ell}(\phi_{\ell})\mathbf{F}\mathbf{F}\herm\mathbf{A}_{\ell}\herm(\phi_{\ell})+\sigma_{\mathrm{B}}^{2}\mathbf{I}$, and $\boldsymbol{\lambda}_{\max}(\mathbf{X})$ denotes the eigenvector corresponding to the largest eigenvalue of the square matrix $\mathbf{X}$. Note that due to the \emph{orthonormality} property of the eigenvectors,~\eqref{eq:UMC-receive-beamforming} is satisfied automatically. 

\subsection{Optimizing $\mathbf{F}$ and $\boldsymbol{\theta}$}

Since the objective in~$(\mathbb{P}1)$ is already convex, we start tackling the constraints~\eqref{eq:P1-comm-QoS}, \eqref{eq:P1-energy-harvesting-QoS}, \eqref{eq:P1-sensing-QoS}, and~\eqref{eq:P1-UMC} one-by-one, with $\mathbf{U}$ being fixed. 

For a given $k\in\mathcal{K}_{\mathrm{I}}$,~\eqref{eq:P1-comm-QoS} can be rewritten as 
\begin{subequations}
\label{eq:comm-SINR-1}
\begin{align}
\frac{1}{\Gamma_{\mathrm{I}k}}|\mathbf{z}_{\mathrm{I}k}\mathbf{f}_{L+k}|^{2}\geq & \ \sigma_{\mathrm{I}k}^{2}+\sum\nolimits_{\imath\in\mathcal{Q}\setminus\{L+k\}}\big(\zeta_{k\imath}^{2}+\bar{\zeta}_{k\imath}^{2}\big),\label{eq:comm-SINR-1-1}\\
\zeta_{k\imath}\geq & \ |\Re\{\mathbf{z}_{\mathrm{I}k}\mathbf{f}_{\imath}\}|,\ \forall\imath\in\mathcal{Q}\setminus\{L+k\},\label{eq:comm-SINR-1-2}\\
\bar{\zeta}_{k\imath}\geq & \ |\Im\{\mathbf{z}_{\mathrm{I}k}\mathbf{f}_{\imath}\}|,\ \forall\imath\in\mathcal{Q}\setminus\{L+k\},\label{eq:comm-SINR-1-3}
\end{align}
\end{subequations}
where $\zeta_{k\imath}$ and $\bar{\zeta}_{k\imath}$ are slack variables. It can be shown that if~\eqref{eq:P1-comm-QoS} is feasible, then so is~\eqref{eq:comm-SINR-1}, and vice-versa; the above transformation does not change the optimality of~($\mathbb{P}1$). We can note that the left-hand side (LHS) of~\eqref{eq:comm-SINR-1-1} is neither convex nor concave (due to the coupling between $\boldsymbol{\theta}_{\mathsf{R}}/\boldsymbol{\theta}_{\mathsf{T}}$ and $\mathbf{f}_{L+k}$), while the right-hand side (RHS) is convex. Therefore, we derive a concave lower bound on $|\mathbf{z}_{\mathrm{I}k}\mathbf{f}_{L+k}|^{2}$. This can be achieved using linearization via first-order Taylor approximation as follows: 
\begin{align}
 & |\mathbf{z}_{\mathrm{I}k}\mathbf{f}_{L+k}|^{2}\geq2\Re\big\{\varphi_{\mathrm{I}k}^{(\varepsilon)}\ \!\!\herm\mathbf{z}_{\mathrm{I}k}\mathbf{f}_{L+k}\big\}-|\varphi_{\mathrm{I}k}^{(\varepsilon)}|^{2}\nonumber \\
=\  & \frac{1}{2}\big\{\|\varphi_{\mathrm{I}k}^{(\varepsilon)}\mathbf{z}_{\mathrm{I}k}\herm+\mathbf{f}_{L+k}\|^{2}-\|\varphi_{\mathrm{I}k}^{(\varepsilon)}\mathbf{z}_{\mathrm{I}k}\herm-\mathbf{f}_{L+k}\|^{2}\big\}-|\varphi_{\mathrm{I}k}^{(\varepsilon)}|^{2}\nonumber \\
\geq\  & \Re\big\{\boldsymbol{\xi}_{\mathrm{I}k}^{(\varepsilon)}\ \!\!\herm\big(\varphi_{\mathrm{I}k}^{(\varepsilon)}\mathbf{z}_{\mathrm{I}k}\herm+\mathbf{f}_{L+k}\big)\big\}-\frac{1}{2}\|\boldsymbol{\xi}_{\mathrm{I}k}^{(\varepsilon)}\|^{2}\nonumber \\
 & -\frac{1}{2}\|\varphi_{\mathrm{I}k}^{(\varepsilon)}\mathbf{z}_{\mathrm{I}k}\herm-\mathbf{f}_{L+k}\|^{2}-|\varphi_{\mathrm{I}k}^{(\varepsilon)}|^{2}\triangleq\mathscr{G}(\mathbf{f}_{L+k},\boldsymbol{\theta}_{\mathrm{I}k}),\label{eq:first_transformation}
\end{align}
where the inequalities follow from the relation $\|\mathbf{x}\|^{2}\geq2\Re\big\{\mathbf{y}\herm\mathbf{x}\big\}-\|\mathbf{y}\|^{2}$, and the equality follows from the relation $\Re\big\{\mathbf{x}\herm\mathbf{y}\big\}=\frac{1}{4}\big(\|\mathbf{x}+\mathbf{y}\|^{2}-\|\mathbf{x}-\mathbf{y}\|^{2}\big)$. Moreover, in~\eqref{eq:first_transformation}, $\varphi_{\mathrm{I}k}^{(\varepsilon)}\triangleq\mathbf{z}_{\mathrm{I}k}^{(\varepsilon)}\mathbf{f}_{L+k}^{(\varepsilon)}$, $\boldsymbol{\xi}_{\mathrm{I}k}^{(\varepsilon)}\triangleq\varphi_{\mathrm{I}k}^{(\varepsilon)}\mathbf{z}_{\mathrm{I}k}^{(\varepsilon)}\ \!\!\herm+\mathbf{f}_{L+k}^{(\varepsilon)}$, and $\mathbf{z}_{\mathrm{I}k}^{(\varepsilon)}=\mathbf{h}_{\mathrm{D}k}+\mathbf{h}_{\mathrm{R}k}\boldsymbol{\Theta}_{\mathrm{I}k}^{(\varepsilon)}\mathbf{H}$. Since~\eqref{eq:first_transformation} is jointly concave with respect to $\boldsymbol{\theta}$ and $\mathbf{F}$, we can rewrite~\eqref{eq:comm-SINR-1-1} equivalently as 
\begin{equation}
\frac{1}{\Gamma_{\mathrm{I}k}}\mathscr{G}(\mathbf{f}_{L+k},\boldsymbol{\theta}_{\mathrm{I}k})\geq\sigma_{\mathrm{I}k}^{2}+\sum\nolimits_{\imath\in\mathcal{Q}\setminus\{L+k\}}\big(\zeta_{k\imath}^{2}+\bar{\zeta}_{k\imath}^{2}\big).\label{eq:comm-SINR-num-final}
\end{equation}
Next, using the fact that $x\geq|y|$ leads to $x\geq\pm y$,~\eqref{eq:comm-SINR-1-2} and~\eqref{eq:comm-SINR-1-3} can be transformed to the following set of inequalities via first-order Taylor approximation for a given $(k,\imath)$:
\begin{align}
\zeta_{k\imath}\geq\psi\big(\pm\mathbf{f}_{\imath},\boldsymbol{\theta}_{\mathrm{I}k}\big),\qquad & \bar{\zeta}_{k\imath}\geq\bar{\psi}\big(\pm j\mathbf{f}_{\imath},\boldsymbol{\theta}_{\mathrm{I}k}\big),\label{eq:comm-SINR-RHS-final}
\end{align}
where $\psi(\mathbf{x},\mathbf{y})\triangleq\frac{1}{4}\big[\|\mathbf{y}\herm+\mathbf{x}\|^{2}-2\Re\big\{\big(\mathbf{y}^{(\varepsilon)}-\mathbf{x}^{(\varepsilon)}\ \!\!\herm\big)\big(\mathbf{y}\herm-\mathbf{x}\big)\big\}+\|\mathbf{y}^{(\varepsilon)}\ \!\!\herm-\mathbf{x}^{(\varepsilon)}\|^{2}\big]$, and $\bar{\psi}(j\mathbf{x},\mathbf{y})\triangleq\frac{1}{4}\big[\|\mathbf{y}\herm-j\mathbf{x}\|^{2}-2\Re\big\{\big(\mathbf{y}^{(\varepsilon)}-j\mathbf{x}^{(\varepsilon)}\ \!\!\herm\big)\big(\mathbf{y}\herm+j\mathbf{x}\big)\big\}+\|\mathbf{y}^{(\varepsilon)}\ \!\!\herm+j\mathbf{x}^{(\varepsilon)}\|^{2}\big]$. Moreover, for the transformation in~\eqref{eq:comm-SINR-RHS-final}, we have used the well-known relations $\Re\big\{\mathbf{x}\herm\mathbf{y}\big\}=\frac{1}{4}\big(\|\mathbf{x}+\mathbf{y}\|^{2}-\|\mathbf{x}-\mathbf{y}\|^{2}\big)$ and $\Im\big\{\mathbf{x}\herm\mathbf{y}\big\}=\frac{1}{4}\big(\|\mathbf{x}-j\mathbf{y}\|^{2}-\|\mathbf{x}+j\mathbf{y}\|^{2}\big)$; the convexification of the negative quadratic term is performed using the first-order Taylor approximation. 

Next for a given $\jmath\in\mathcal{K}_{\mathrm{E}}$, we note that the LHS of the constraint in~\eqref{eq:P1-energy-harvesting-QoS} is non-concave. This calls for obtaining a concave lower bound for $\frac{1}{\Delta_{\jmath}}p_{\mathrm{E}\jmath}(\mathbf{F},\boldsymbol{\theta})$. One can obtained this following a procedure similar to~\eqref{eq:first_transformation}, yielding to the following reformulation of~\eqref{eq:P1-energy-harvesting-QoS}:
\begin{equation}
\frac{\eta}{\Delta_{\jmath}}\sum\nolimits_{\imath\in\mathcal{Q}}\mathscr{G}(\mathbf{f}_{\imath},\boldsymbol{\theta}_{\mathrm{E}\jmath})\geq1.\label{eq:energy-harvesting-QoS-final}
\end{equation}

We now turn our focus on the sensing constraints in~\eqref{eq:P1-sensing-QoS}. For a given $l\in\mathcal{L}$, and fixed $\mathbf{u}_{l}$, the constraint can be rewritten as follows:
\begin{align}
 & \frac{\bar{\alpha}}{\Gamma_{\mathrm{T}l}}\sum\nolimits_{\imath\in\mathcal{Q}}|\mathbf{u}_{l}\herm\mathbf{A}_{l}(\phi_{l})\mathbf{f}_{\imath}|^{2}\geq\sum\nolimits_{\imath\in\mathcal{Q}}|\mathbf{u}_{l}\herm\mathbf{G}\mathbf{f}_{\imath}|^{2}\nonumber \\
 & \ \ +\sum\nolimits_{\ell\in\mathcal{L}\setminus\{l\}}\bar{\alpha}_{\ell}\sum\nolimits_{\imath\in\mathcal{Q}}|\mathbf{u}_{l}\herm\mathbf{A}_{\ell}(\phi_{\ell})\mathbf{f}_{\imath}|^{2}+\sigma_{\mathrm{B}}^{2}\|\mathbf{u}_{l}\|^{2}.\label{eq:sensing-QoS-1}
\end{align}
Note that both the sides of~\eqref{eq:sensing-QoS-1} are convex, and therefore, we linearize the LHS, yielding an equivalent reformulation of~\eqref{eq:P1-sensing-QoS} given in~\eqref{eq:sensing-QoS-final}, shown on the next page, where $\varpi_{l\imath}^{(\varepsilon)}\triangleq\mathbf{u}_{l}\herm\mathbf{A}_{l}(\phi_{l})\mathbf{f}_{\imath}^{(\varepsilon)}$. 
\begin{figure*}
\begin{centering}
\begin{equation}
\frac{\bar{\alpha}}{\Gamma_{\mathrm{T}l}}\sum\nolimits_{\imath\in\mathcal{Q}}\Big[2\Re\big\{\varpi_{l\imath}^{(\varepsilon)}\ \!\!\herm\mathbf{u}_{l}\herm\mathbf{A}_{l}(\phi_{l})\mathbf{f}_{\imath}\big\}-|\varpi_{l\imath}|^{2}\Big]\geq\sum\nolimits_{\imath\in\mathcal{Q}}|\mathbf{u}_{l}\herm\mathbf{G}\mathbf{f}_{\imath}|^{2}+\sum\nolimits_{\ell\in\mathcal{L}\setminus\{l\}}\bar{\alpha}_{\ell}\sum\nolimits_{\imath\in\mathcal{Q}}|\mathbf{u}_{l}\herm\mathbf{A}_{\ell}(\phi_{\ell})\mathbf{f}_{\imath}|^{2}+\sigma_{\mathrm{B}}^{2}\|\mathbf{u}_{l}\|^{2}.\label{eq:sensing-QoS-final}
\end{equation}
\hrulefill
\par\end{centering}
\end{figure*}

The only non-convexity remaining now in~($\mathbb{P}1$) is due to the non-convex constraint in~\eqref{eq:P1-UMC}. To tackle this, we first relax the equality in~\eqref{eq:P1-UMC} to be an convex inequality, and then force the inequality to be the equality by including a regularization term in the objective function. This transformation yields the following equivalent formulation of~($\mathbb{P}1$): 
\begin{subequations}
\label{eq:P2}
\begin{eqnarray}
\!\!\!\!\!\!\!\!\!\!\!\!(\mathbb{P}2) & \underset{\mathbf{F},\boldsymbol{\theta},\boldsymbol{\zeta},\bar{\boldsymbol{\zeta}}}{\minimize} & \|\mathbf{F}\|^{2}-\rho\big[2\Re\big\{\boldsymbol{\theta}^{(\varepsilon)}\ \!\!\herm\boldsymbol{\theta}\big\}-\|\boldsymbol{\theta}^{(\varepsilon)}\|^{2}\big],\label{eq:P2-objective}\\
 & \st & \eqref{eq:comm-SINR-num-final}\ \forall k\in\mathcal{K}_{\mathrm{I}},\label{eq:P2-comm-SINR-num}\\
 &  & \eqref{eq:comm-SINR-RHS-final}\ \forall k\in\mathcal{K}_{\mathrm{I}},\imath\in\mathcal{Q}\setminus\{L+k\},\\
 &  & \eqref{eq:energy-harvesting-QoS-final}\ \forall\jmath\in\mathcal{K}_{\mathrm{E}},\\
 &  & \eqref{eq:sensing-QoS-final}\ \forall l\in\mathcal{L},\\
 &  & |\theta_{\mathsf{R},\varkappa}|^{2}+|\theta_{\mathsf{T},\varkappa}|^{2}\leq1\ \forall\varkappa\in\mathscr{N}_{\mathrm{s}}.
\end{eqnarray}
\end{subequations}
Here $\boldsymbol{\zeta}\triangleq\{\zeta_{k\imath}\}_{k\in\mathcal{K}_{\mathrm{I}},\imath\in\mathcal{Q}\setminus\{L+k\}}$, $\bar{\boldsymbol{\zeta}}\triangleq\{\bar{\zeta}_{k\imath}\}_{k\in\mathcal{K}_{\mathrm{I}},\imath\in\mathcal{Q}\setminus\{L+k\}}$, and $\rho>0$ is the regularization parameter. It is not difficult to show that~($\mathbb{P}2$) is a second-order cone programming (SOCP) problem that can be efficiently solved using off-the-shelf solvers. We summarize the proposed AO-based solution to~($\mathbb{P}1$) in \textbf{Algorithm~\ref{algo}}.

\subsection{Convergence Analysis}

Note that since the objective function in~$(\mathbb{P}1)$ is not a function of $\mathbf{U}$, the convergence of the AO-based method in \textbf{Algorithm~\ref{algo}} is dominated by the convergence behavior of the solution to~$(\mathbb{P}2)$. Define the objective in~$(\mathbb{P}2)$, before linearizing the regularization term, by $\mathscr{O}(\mathbf{F},\boldsymbol{\theta})\triangleq\|\mathbf{F}\|^{2}-\rho\|\boldsymbol{\theta}\|^{2}$. Then for a given value of the regularization parameter $\rho$, the optimal solution to~$(\mathbb{P}2)$ in the $\varepsilon$-th iteration is also a feasible solution in the $\varepsilon+1$-th iteration. Therefore, $\mathscr{O}\big(\mathbf{F}^{(\varepsilon+1)},\boldsymbol{\theta}^{(\varepsilon+1)}\big)\leq\mathscr{O}\big(\mathbf{F}^{(\varepsilon)},\boldsymbol{\theta}^{(\varepsilon)}\big)$, i.e., the algorithm generates a \emph{non-increasing sequence} of the regularized objective. Since the objective $\mathscr{O}(\mathbf{F},\boldsymbol{\theta})$ is bounded from the below due to the constraint in~$(\mathbb{P}2)$, i.e., $-\rho N_{\mathrm{s}}\leq\mathscr{O}(\mathbf{F},\boldsymbol{\theta})$, it can directly be inferred that $\mathscr{O}(\mathbf{F},\boldsymbol{\theta})$ is convergent. The proof for the sequence of iterates being convergent, and the limit point being a stationary solution is rather standard, and therefore omitted for the sake of brevity. 

\subsection{Analysis of Computational Complexity}

The per-iteration complexity of \textbf{Algorithm~\ref{algo}} is dominated by the operations in lines 3 and 4 of the algorithm. The complexity for computing $\mathbf{U}_{\mathrm{opt}}$ is dominated by the matrix multiplication operations, matrix inversion, and eigenvalue decomposition operations in~\eqref{eq:optimal-receive-beamforming-closed}, which entails a complexity of $\mathcal{O}(L(3N_{\mathrm{r}}^{3}+4N_{\mathrm{t}}^{2}N_{\mathrm{r}}+2N_{\mathrm{t}}N_{\mathrm{r}}^{2}))$. Next, by computing the number of quadratic conic constraints, and the size of each of those constraints in~$(\mathbb{P}2)$, one can show that the per-iteration complexity of line~4 in \textbf{Algorithm~\ref{algo}} is well-approximated by $\mathcal{O}(N_{\mathrm{s}}^{3.5})$~\cite[Sec. 6.6.2]{ModernOptLectures}. Since, in a practical setting, $N_{\mathrm{s}}\gg\max\{N_{\mathrm{t}},N_{\mathrm{r}},K_{\mathrm{I}},K_{\mathrm{E}},L\}$, the overall per-iteration complexity of \textbf{Algorithm~\ref{algo}} can be approximated by $\mathcal{O}(N_{\mathrm{s}}^{3.5})$. 

\section{Results and Discussion}

\begin{figure}[tb]
\begin{centering}
\includegraphics[width=0.75\columnwidth]{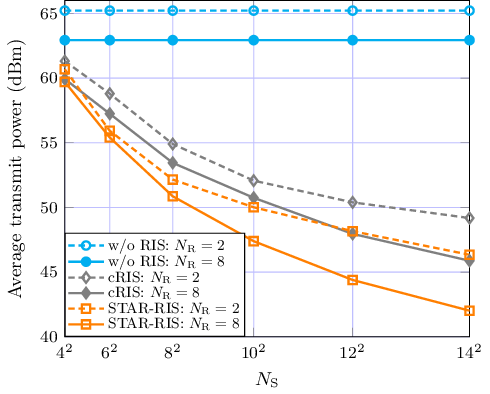}
\par\end{centering}
\caption{Impact of varying $N_{\mathrm{s}}$ on the required transmit power for $N_{\mathrm{t}}=6$, $\Gamma_{\mathrm{I}}=10$~dB, $\Gamma_{\mathrm{T}}=5$~dB, and $\Delta=0.1$~mW.}
\label{fig_vary_nS}
\end{figure}

\begin{figure}[tb]
\begin{centering}
\includegraphics[width=0.7\columnwidth]{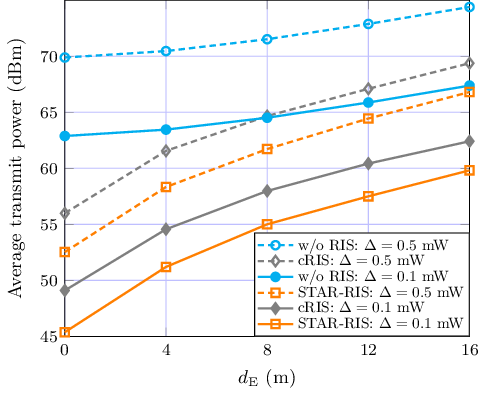}
\par\end{centering}
\caption{Impact of increasing the ERs distance on the required transmit power for $N_{\mathrm{t}}=6$, $N_{\mathrm{r}}=8$, $N_{\mathrm{s}}=128$, $\Gamma_{\mathrm{I}}=10$~dB, and $\Gamma_{\mathrm{T}}=5$~dB.}
\label{fig:vary_ER_distance}
\end{figure}

\begin{figure}[tb]
\begin{centering}
\includegraphics[width=0.7\columnwidth]{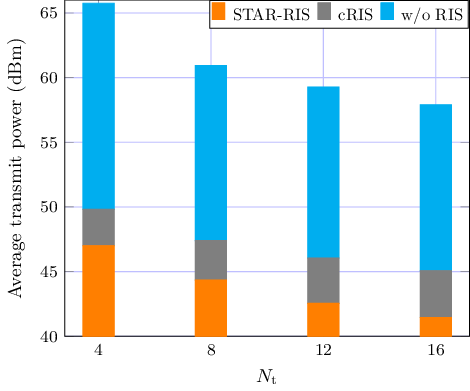}
\par\end{centering}
\caption{Impact of varying $N_{\mathrm{t}}$ on the required transmit power for $N_{\mathrm{r}}=8$, $N_{\mathrm{s}}=128$, $\Gamma_{\mathrm{I}}=10$~dB, $\Gamma_{\mathrm{T}}=5$~dB, and $\Delta=0.1$~mW.}
\label{fig:vary_nT}
\end{figure}

In this section, we present extensive numerical experiments to evaluate the transmit power requirements of the STAR-RIS-ISCPT system. We assume that B is located at $(0,0,2.5)$~m, and S is located at $(20,2,2.5)$~m. The IRs and ERs in the reflection region are randomly located inside the circle of radius $1$~m, centered at $(20,-50,0)$~m and $(20,0,0)$~m, respectively, while those in the transmission region are randomly located inside the circle of radius $1$~m, centered at $(20,50,0)$~m and $(20,4,0)$~m, respectively. We model $\mathbf{H}$, $\mathbf{h}_{\mathrm{R}k}\ (\forall k\in\mathcal{K}_{\mathrm{I}})$, and $\mathbf{g}_{\mathrm{R}\jmath}\ (\forall\jmath\in\mathcal{K}_{\mathrm{E}})$ following a Rician distribution with a Rice factor of $3$~dB, and path loss exponent of 2.2. At the same time, $\mathbf{h}_{\mathrm{D}k}\ (\forall k\in\mathcal{K}_{\mathrm{I}})$ and $\mathbf{g}_{\mathrm{D}\jmath}\ (\forall\jmath\in\mathcal{K}_{\mathrm{E}})$ are modeled using Rayleigh distribution with path loss exponent of 3.6. The self-interference link $\mathbf{G}$ is modeled as a Rayleigh fading channel with a variance of $0$~dB with respect to the noise power. We consider two passive targets at distances 20~m and $30$~m from B with azimuth angles of $-30^{\circ}$ and $-60^{\circ}$, respectively. The path loss between two nodes at distance $d$~m is modeled as $(-30-10\beta\log_{10}(d/d_{0}))$~dB with $d_{0}=1$~m. Without loss of generality, we assume $K_{\mathrm{I}}=K_{\mathrm{E}}=L=2$, $\Gamma_{\mathrm{I}k}=\Gamma_{\mathrm{I}}\ (\forall k\in\mathcal{K}_{\mathrm{I}})$, $\Delta_{\jmath}=\Delta\ (\forall\jmath\in\mathcal{K}_{\mathrm{E}})$, $\Gamma_{\mathrm{T}l}=\Gamma_{\mathrm{T}}\ (\forall l\in\mathcal{L})$, and $\sigma_{\mathrm{I}k}^{2}=\sigma_{\mathrm{B}}^{2}=-90$~dB $(\forall k\in\mathcal{K}_{\mathrm{I}})$. The average transmit power in~Figs.~\ref{fig_vary_nS}–\ref{fig:vary_nT} are computed over 100 independent set of channel realizations. The regularization parameter $\rho$ is set to $0.01$, $\eta=0.8$, and $\bar{\alpha}=0.5$. For comparing the performance of the proposed STAR-RIS-enabled ISCPT system, we consider a conventional RIS (cRIS)-enabled ISCPT, and an ISCPT system without RIS (w/o RIS) as benchmarks. In the case of cRIS-enabled system, we assume that the cRIS is equipped with a purely reflection-type RIS with $N_{\mathrm{s}}/2$ elements and a purely transmission-type RIS with $N_{\mathrm{s}}/2$ elements, resulting in the total $N_{\mathrm{s}}$ cRIS elements. We also assume the there are equal number of IRs (respectively ERs) in the reflection and transmission regions of the STAR-RIS- / cRIS-enabled system.
\begin{algorithm}[tb]
\caption{Proposed AO-based Solution to~$(\mathbb{P}1)$.}

\label{algo}

\KwIn{$\mathbf{F}^{(0)}$, $\boldsymbol{\theta}^{(0)}$}

$\varepsilon\leftarrow0$\;

\Repeat{convergence }{

For the given $\mathbf{F}^{(\varepsilon)}$ and $\boldsymbol{\theta}^{(\varepsilon)}$, obtain $\mathbf{U}_{\mathrm{opt}}$ using~\eqref{eq:optimal-receive-beamforming-closed}\;

For the given $\mathbf{U}_{\mathrm{opt}}$, solve~$(\mathbb{P}2)$ and denote the obtained solution as $\mathbf{F}_{\mathrm{opt}}$, $\boldsymbol{\theta}_{\mathrm{opt}}$\;

Update: $\mathbf{F}^{(\varepsilon+1)}\leftarrow\mathbf{F}_{\mathrm{opt}}$, $\boldsymbol{\theta}^{(\varepsilon+1)}\leftarrow\boldsymbol{\theta}_{\mathrm{opt}}$\;

$\varepsilon\leftarrow\varepsilon+1$\;

}

\KwOut{$\mathbf{F}_{\mathrm{opt}}$, $\boldsymbol{\theta}_{\mathrm{opt}}$, $\mathbf{U}_{\mathrm{opt}}$}
\end{algorithm}

\begin{figure}[tb]
\begin{centering}
\includegraphics[width=0.76\columnwidth]{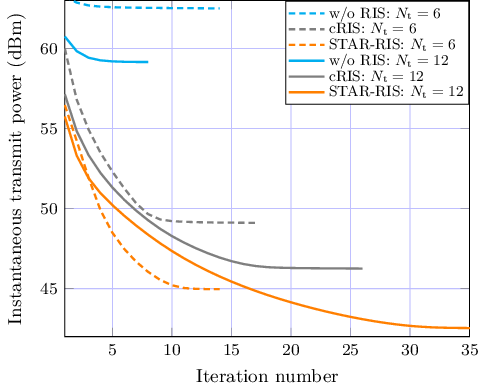}
\par\end{centering}
\caption{Convergence of the proposed algorithm for $N_{\mathrm{r}}=8$, $N_{\mathrm{s}}=128$, $\Gamma_{\mathrm{I}}=10$~dB, $\Gamma_{\mathrm{T}}=5$~dB, and $\Delta=0.1$~mW.}
\label{fig:convergence_sequence}
\end{figure}

In Fig.~\ref{fig:convergence_sequence}, we show the convergence behavior of the proposed algorithm for $N_{\mathrm{r}}=8$, $N_{\mathrm{s}}=128$, $\Gamma_{\mathrm{I}}=10$~dB, $\Gamma_{\mathrm{T}}=5$~dB, and $\Delta=0.1$~mW. It is evident from the figure that the iterates decrease monotonically, and due to the existence of a lower-bound on the required transmit power to satisfy the QoS constraints, convergence is guaranteed to be achieved. It is interesting to note that on one hand, the required transmit power do not improve much compared to the initial value, in the case of w/o RIS system since the active and receive beamforming vectors are the only optimization variables in the system, leading to a very limited degree-of-freedoms (DoFs). On the other hand, for the case of STAR-RIS- / cRIS-enabled systems, the required transmit power decreases $(\approx)$10~dB from the initial value, since these system have active, receive, and passive beamforming vectors as optimization variables, leading to higher DoFs. 

Fig.~\ref{fig_vary_nS} showcases the impact the both the number of metasurface elements ($N_{\mathrm{s}}$) and the number of receive antennas ($N_{\mathrm{r}}$) at B on the system's transmit power requirement for $N_{\mathrm{t}}=6$, $\Gamma_{\mathrm{I}}=10$~dB, $\Gamma_{\mathrm{T}}=5$~dB, and $\Delta=0.1$~mW. One can note from the figure that increasing $N_{\mathrm{s}}$ has a highly beneficial effect on both cRIS- and STAR-RIS-enabled systems compared to the w/o RIS counterpart in terms of transmit power requirement. For example, under the given system setting, a large-sized cRIS/STAR-RIS (say with $14^{2}$ elements) brings down the transmit power requirement from highly impractical values ($65$~dBm for $N_{\mathrm{r}}=2$ in the w/o RIS system) to practical values (less than $50$~dBm for $N_{\mathrm{r}}=2$). This performance benefit comes from the beamforming gain of the metasurface elements in the cRIS/STAR-RIS system. The experiment also confirms the superiority of STAR-RIS over cRIS, leading to nearly $2.8$~dB improvement for $(N_{\mathrm{s}},N_{\mathrm{r}})=(14^{2},2)$, and $3.8$~dB improvement for $(N_{\mathrm{s}},N_{\mathrm{r}})=(14^{2},8)$. At the same time, increasing the number of receive antennas at B also has notable beneficial impact on the transmit power requirement. As $N_{\mathrm{r}}$ increases, the receive beamforming gain at B for target sensing increases. This in-turn reduces the power requirement to satisfy the sensing QoSs in~\eqref{eq:P1-sensing-QoS}, resulting in the overall reduction in the total power requirement of the system. 

In Fig.~\ref{fig:vary_ER_distance}, we show the impact of ERs' distance on the transmit power requirement of the ISCPT system. Here we assume that the ERs are located inside a circle of radius $1$~m, centered at $(20,-d_{\mathrm{E}},0)$~m in the reflection region, and $(20,4+d_{\mathrm{E}},0)$~m in the transmission region. As the average distance of the ERs from the base station and STAR-RIS/cRIS increases, the path loss of the ERs' link also increases. This in turn reduces the power of the incident signals at the ERs, and therefore, a higher transmit power is required from B to satisfy~\eqref{eq:P1-energy-harvesting-QoS}. The results also indicates that for a given transmit power budget from the base station B, the STAR-RIS-enabled system extends the operating distance of the ERs. 

We show the impact of the number of transmit antennas ($N_{\mathrm{t}}$) at B in Fig.~\ref{fig:vary_nT}, for $N_{\mathrm{r}}=8$, $N_{\mathrm{s}}=128$, $\Gamma_{\mathrm{I}}=10$~dB, $\Gamma_{\mathrm{T}}=5$~dB, and $\Delta=0.1$~mW. By increasing $N_{\mathrm{t}}$, the active beamforming gain increases for the ISCPT system, resulting in a reduction in the minimum required transmit power. This effect is clearly evident from the figure, and the superiority of the STAR-RIS-enabled system over the other benchmark systems is also reconfirmed. Our numerical experiments confirm that the performance difference between the STAR-RIS and cRIS systems varies in the range of $2.8$~dB to $3.6$~dB, while that between STAR-RIS and w/o RIS system varies in the range of $16.4$~dB to $18.8$~dB, highlighting the suitability of STAR-RIS as an excellent enabler for next-generation green wireless.

\section{Conclusion}

In this paper, we have considered the problem of transmit power minimization for a STAR-RIS-enabled ISCPT system. For this purpose, we have formulated a non-convex optimization problem to jointly obtain the active, passive, and receive beamforming design. To tackle the issue of variable coupling in the formulated problem, and to obtain the optimal beamforming design, we used an AO-based approach, where the optimal receive beamforming was obtained in closed-form, while the active and passive beamformers were obtained using SOCP. With the help of numerical experiments, we have shown that STAR-RIS significantly reduce the transmit power requirement for the given setup, and is also significantly beneficial in extending the operating distance of energy receivers for a given transmit power budget, as compared to the cRIS and w/o RIS counterparts. 

\bibliographystyle{IEEEtran}
\bibliography{references}

\end{document}